\documentclass{emulateapj}
\usepackage{apjfonts,amsmath}
\usepackage{color}

\newcommand{\form}{\mbox{H$_2$CO}}

\newcommand{\pcc}{~\mbox{cm$^{-3}$}}

\newcommand{\nh}{\mbox{$n$({\rm H}$_2$)}}

\newcommand{\tcmb}{\mbox{$T_{\rm CMB}$}}
\newcommand{\tkin}{\mbox{$T_{\rm kin}$}}
\newcommand{\tx}{\mbox{$T_{{\rm x}}$}}

\newcommand{\ten}[1]{\mbox{$10^{#1}$}}

\shorttitle{Formaldehyde Silhouettes Against the Cosmic Microwave Background}
\shortauthors{Darling \& Zeiger}

\begin{document}
\title{Formaldehyde Silhouettes Against the Cosmic Microwave Background:  A Mass-Limited, Distance-Independent, 
Extinction-Free Tracer of Star Formation Across the Epoch of Galaxy Evolution
}

\author{ Jeremy Darling\altaffilmark{1,2,3} and Benjamin Zeiger\altaffilmark{1,4} }
\altaffiltext{1}{Center for Astrophysics and Space Astronomy,
Department of Astrophysical and Planetary Sciences,
University of Colorado, 389 UCB, Boulder, CO 80309-0389}
\altaffiltext{2}{NASA Lunar Science Institute, NASA Ames Research Center, Moffett Field, CA}
\altaffiltext{3}{jdarling@colorado.edu}
\altaffiltext{4}{benjamin.zeiger@colorado.edu}

\begin{abstract}
We examine the absorption of cosmic microwave background (CMB) photons
by formaldehyde (\form) over cosmic time.  The $K$-doublet rotational transitions 
of \form\ become ``refrigerated'' --- their excitation temperatures
are driven below the CMB temperature --- via
collisional pumping by molecular hydrogen (H$_2$).  
``Anti-inverted'' \form\ line ratios thus provide
an accurate measurement of the H$_2$ density in molecular clouds.
Using a radiative transfer model, we demonstrate that \form\ centimeter
wavelength line excitation
and detectability are nearly independent of redshift or gas kinetic temperature.
Since the \form\ $K$-doublet lines absorb CMB light, and since the CMB
lies behind every galaxy and provides an exceptionally uniform extended
illumination source, 
\form\ is a distance-independent, extinction-free molecular gas mass-limited 
tracer of dense gas in galaxies.  A Formaldehyde Deep Field 
could map the history of cosmic star formation in a uniquely unbiased
fashion and 
may be possible with large bandwidth wide-field radio interferometers whereby the
silhouettes of star-forming galaxies would be detected across
the epoch of galaxy evolution.
We also examine the possibility that \form\ lines may provide a standardizable 
galaxy ruler for cosmology similar to the Sunyaev-Zel'dovich effect in galaxy clusters
but applicable to much higher redshifts and larger samples.
Finally, we explore how anti-inverted meterwave \form\ lines 
in galaxies during the peak of cosmic star formation
may contaminate \ion{H}{1} 21 cm tomography of the Epoch of Reionization.
\end{abstract}

\section{Introduction} 
Centimeter-wave $K$-doublet rotational
transitions of formaldehyde (\form; Figure \ref{H2CO_levels}) 
can be collisionally ``refrigerated'' below the cosmic microwave
background (CMB) temperature such that these transitions absorb CMB photons
\citep{palmer69,townes69,evans75,garrison75}.  The manner in which the centimeter
lines of \form\ are cooled is equivalent to a maser pumping process: a
pump drives an over-population of states, but in this case they are 
lower-energy states, and the level populations
become ``anti-inverted'' compared to thermal.  When the line excitation
temperature drops below the local CMB temperature, CMB photons may be absorbed.
Since absorption lines are detectable independent of distance, and since the CMB 
illuminates all gas and provides an exceptionally uniform
illumination that lies behind every galaxy, a  Formaldehyde Deep Field (FDF) can
provide a {\it mass-limited} census of the dense molecular gas
associated with star formation.  If multiple lines
are observed then physical molecular gas densities may also be obtained fairly independently of 
the gas kinetic temperature or other factors \citep[e.g.,][]{mangum08}.

In this Letter, we examine the properties of the anti-inverted centimeter-wave
\form\ $K$-doublet lines as 
a function of physical conditions and cosmological redshift 
in a feasibility study for a census of the cosmic gas 
evolution and star formation history via an FDF.  
We also discuss the possibility that \form\ lines can provide a 
standardizable ruler for cosmology and may contaminate \ion{H}{1} 21 cm 
tomography observations of the Epoch of Reionization (EoR).
This work examines radiative transfer
models only, but is supported by observations of
our galaxy \citep{mangum93,ginsburg11}, nearby starburst galaxies 
\citep{mangum08}, and the 
gravitational lens galaxy toward B0218+357 at $z=0.68$
\citep{zeiger10}.  

We assume a no-curvature cosmology with 
$H_\circ = 72$~km~s$^{-1}$~Mpc$^{-1}$,
$\Omega_m = 0.26$, and  $\Omega_\Lambda = 0.74$.  

\section{Radiative Transfer}

\begin{figure}
\epsscale{1.15}
\plotone{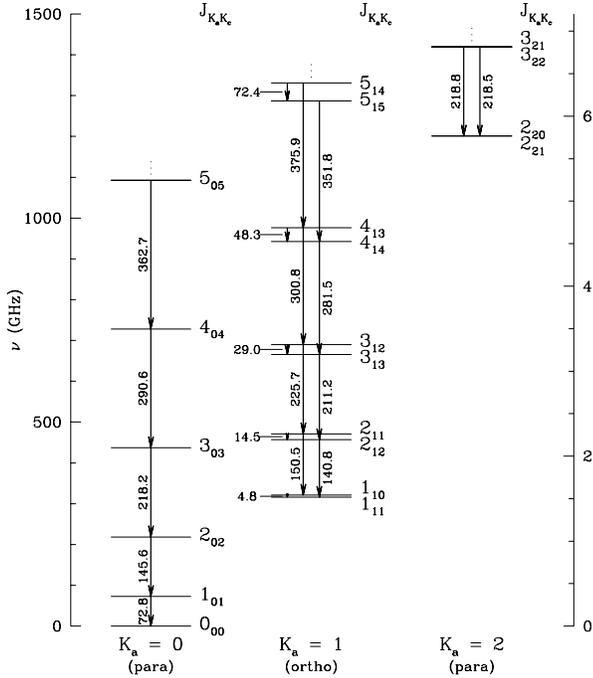}
\caption{Formaldehyde energy levels for the first three
rotation ladders.  Arrows indicate allowed transitions,
and line frequencies are listed in GHz.
The $K$-doublet levels splitting $J$
states of the $K_a=1$ ortho-\form\ rotation ladder are the focus of this Letter.  The wavelengths
of these transitions are, from lowest energy to highest,
roughly 6, 2, 1, 0.6, and 0.4~cm.
\label{H2CO_levels}}
\end{figure}

We use the RADEX one-dimensional non-LTE radiative transfer code 
with the escape probability determined by the large velocity gradient 
\citep[LVG;][]{sobolev60,goldreich74}
approximation to examine 
the excitation and thus detectability and interpretation of 
cm \form\ lines \citep{vandertak07}.   
We employ the H$_2$ collision rate coefficients calculated by \citet{troscompt09b} for the 10 lowest
levels of ortho-\form\ (to $J=5$; Figure \ref{H2CO_levels})
for gas kinetic temperatures of 5--100~K.
We have compared the results from this code
with another LVG treatment \citep{henkel80} 
and with RADEX (using the LVG approximation), both using the \citet{green91}
collision rates scaled from He to H$_2$, and all are in good agreement for low H$_2$ ortho-to-para ratio (OPR).
The latter models include the first 40 energy levels of ortho-\form; agreement
between these models and those using only the 10 levels of \citet{troscompt09b} 
lends credibility to our treatment of warm molecular gas in  star-forming galaxies.

For the radiative transfer calculations 
we assume statistical equilibrium but allow all line excitation temperatures to float
because the dominant collision partner, H$_2$, ``pumps'' \form\ into 
non-thermal excitation state distributions.
We assume that the CMB is the dominant continuum source and that there
is no significant contribution 
to the radio continuum from the host galaxy itself.  An examination of this
effect in the starburst galaxy M81 shows a small effect that does not 
qualitatively alter our results \citep{mangum08}.  
We also assume an \form/H$_2$ abundance of $10^{-9}$, which is 
reasonable in molecular gas-rich galaxies, even at high redshift 
(Zeiger \& Darling 2010 obtain an abundance 0.1--$2.5\times10^{-9}$ 
in the gravitational lens galaxy toward B0218+357 at $z=0.68$),
and an \form\ OPR of 3. 

\citet{troscompt09a, troscompt09b} show that the anti-inversion
of ortho-\form\ is principally driven by collisions with para-H$_2$.
Since the \form\ line excitation temperatures critically determine the
detectability and interpretation of lines, 
the effect of the unknown H$_2$ OPR is a concern.  
To address this, we examine model runs with H$_2$ OPR of 3 and $10^{-2}$
(effectively zero) and find small differences in line excitation temperature
($\lesssim0.3$~K) and no significant effect on the broad results and interpretation (Section \ref{sec:results}).
The basic phenomenon of anti-inversion 
of cm \form\ lines and their
ability to produce silhouettes of star-forming galaxies against the
CMB is robust to the unknown H$_2$ OPR,
as is the nearly temperature- and redshift-independent 
nature of the line detectability and excitation.

We examine the 
model predictions for cm \form\ line excitation
for varying $n({\rm H}_2)$, gas kinetic temperature, and redshift.
When not varying, we assume $n({\rm H}_2)=10^4$~cm$^{-3}$,
$T_{\rm kin}=40$~K, and H$_2$ OPR $10^{-2}$
(e.g., \citet{maret07}, \citet{pagani09}, and \citet{troscompt09a} infer 
ratios in the range 0.01--0.1 in cold Galactic pre-stellar molecular gas).
We focus on the first five
$K$-doublet lines because these produce the strongest signals
(millimeter lines thermalize in typical conditions), and
the \citet{troscompt09b} calculations do not include higher states.
These lines have wavelengths of roughly 6, 2, 1, 0.6, and 0.4~cm
(4.8, 14.5, 29.0, 48.3, and 72.4~GHz, respectively).

The observed line temperature depends on the decrement
between the rest-frame excitation temperature and CMB temperature,
redshifted to $z=0$, and the line optical depth:
\begin{equation}
 \Delta T_{\rm Obs}={T_{\rm x}(z)-T_{\rm CMB}(z)\over1+z}(1-e^{-\tau})
\label{eqn:DTobs}
\end{equation}
where 
$T_{\rm CMB}(z)=T_{{\rm CMB},0}(1+z)$ and $T_{{\rm CMB},0}=2.73$~K.
For a non-unity covering factor, the optical depth would be 
correspondingly diminished.  
 \citet{mangum08} estimate the apparent line optical depths 
in nearby star-forming galaxies: 0.002--0.007 in the 6~cm line and
0.0015--0.005 in the 2~cm line.   \citet{zeiger10} 
estimate 0.017 at 6~cm and 0.008 at 2~cm 
in the molecular absorber toward B0218+357 at $z=0.67$.
The Galactic range among ultracompact \ion{H}{2}
regions is similar, but with a higher upper bound:  0.006--1.0 at 6~cm
and 0.003--0.3 at 2~cm \citep{ginsburg11}.

The goal of this work is to examine the observer-frame temperature decrement
$(T_{\rm x}(z)-T_{\rm CMB}(z))/(1+z)$ of cm \form\ lines,
which is equivalent to $\Delta T_{\rm Obs}$ for optically thick lines.  
We focus on the 
excitation temperature of the cm \form\ lines as a function
of gas kinetic temperature, H$_2$ density, and redshift because 
these factors will most significantly determine the utility and 
interpretation of \form\ line observations at high redshift.

\begin{figure}
\epsscale{1.25}
\plotone{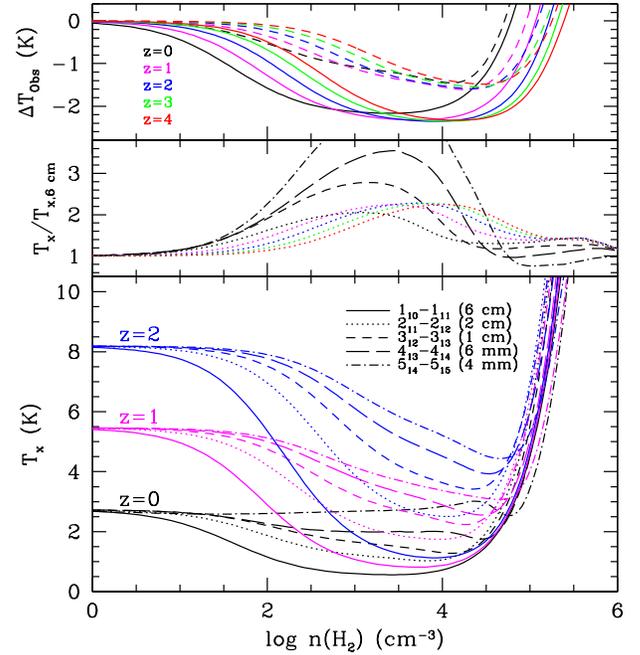}
\caption{Formaldehyde line excitation 
vs.\  H$_2$ number density.
We plot several redshifts to demonstrate the excitation 
redshift dependence.  
Line style indicates transition, and color indicates redshift. 
Bottom: excitation temperature vs.\  $\log$ \nh.  
Middle: excitation temperature normalized by the 6~cm line excitation temperature vs.\  \nh.  
We plot all five lines at $z=0$ and the 2~cm line at redshifts
$z=0$--4, normalized to the 6~cm line at each redshift.  
Top:  observer-frame temperature decrement, $\Delta T_{\rm Obs}$ (Equation \ref{eqn:DTobs}) for
an optically thick line vs.\  \nh.  We plot redshifts $z=0$--4 for
the representative 6~cm and the 1~cm lines (other lines are omitted for clarity).  
$\Delta T_{\rm Obs}=0$~K represents the 
transition from absorption to emission.
\label{Tx_vs_nH2}}
\end{figure}

\section{Results}\label{sec:results}

Our radiative transfer models show
that the excitation temperature of cm \form\ lines is a strong function of H$_2$ number
density, collisionally driven below $T_{\rm CMB}$ 
over roughly three orders of magnitude, 
$10^2$~cm$^{-3}\lesssim n({\rm H}_2) \lesssim 10^5$~cm$^{-3}$
(Figure \ref{Tx_vs_nH2}).
At low densities the \form\ transitions thermalize to the 
CMB temperature, $T_{\rm x} \rightarrow T_{\rm CMB}$,  
because the molecule becomes decoupled from the gas and
assumes the radiation temperature.  
At high densities the \form\ excitation thermalizes to the 
kinetic temperature, $T_{\rm x} \rightarrow T_{\rm kin}$.  

The range over which \nh\ causes anti-inversion of the \form\
lines shifts slightly to higher densities with increasing redshift,
by $\sim$0.5 dex from $z=0$ to $z=4$, but this is entirely due to the $T_{\rm CMB}=(1+z)~2.73$~K 
scaling rather than a change in physical pumping conditions.
The minimum $\Delta T_{\rm Obs}$ 
likewise grows slightly (becomes more negative) with increasing redshift, 
decreasing by $\sim$0.2~K from $z=0$ to $z\geq1$.

\begin{figure}
\epsscale{1.15}
\plotone{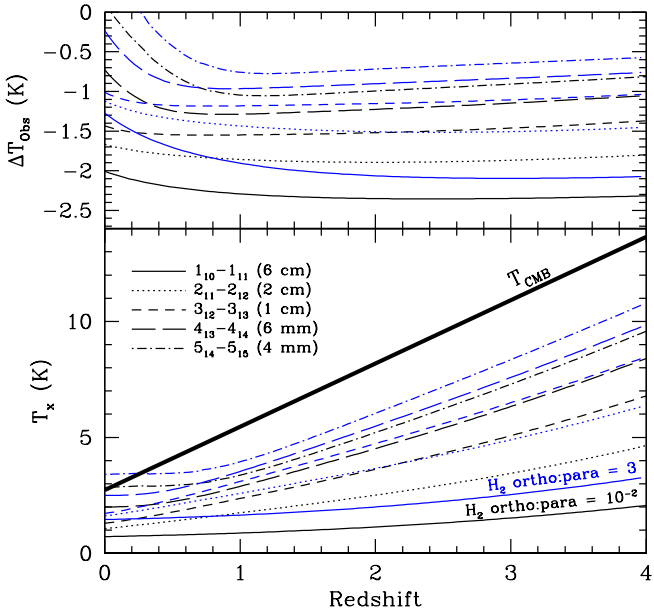}
\caption{Formaldehyde line excitation 
vs.\  redshift.  Line style indicates transition, and color 
indicates H$_2$ OPR.
Bottom:  excitation temperature vs.\  redshift.  
The CMB temperature is bold.  
$T_{\rm x}<T_{\rm CMB}$ indicates absorption of CMB photons.
Top:  observer-frame temperature decrement $\Delta T_{\rm Obs}$ (Equation \ref{eqn:DTobs}) for
an optically thick line vs.\  redshift.  The fairly flat trend in
these decrements shows the nearly redshift-independent
detectability of cm \form\ lines.
\label{Tx_vs_z}}
\end{figure}
 
For fixed \nh\ and \tkin, 
$T_{\rm x}$ grows roughly linearly with redshift (Figure \ref{Tx_vs_z}).  But since the 
CMB temperature scales as $T_{\rm CMB}=(1+z)\,2.73$~K, the redshift dependence
largely drops out of $\Delta T_{\rm Obs}$, and the observable line decrement changes
very weakly with redshift.  Thus, not only is there no distance dimming of 
anti-inverted cm \form\
lines thanks to the absorption of CMB photons, but the excitation conditions are not
changed significantly by the growth of CMB temperature with redshift.

\begin{figure}
\epsscale{1.15}
\plotone{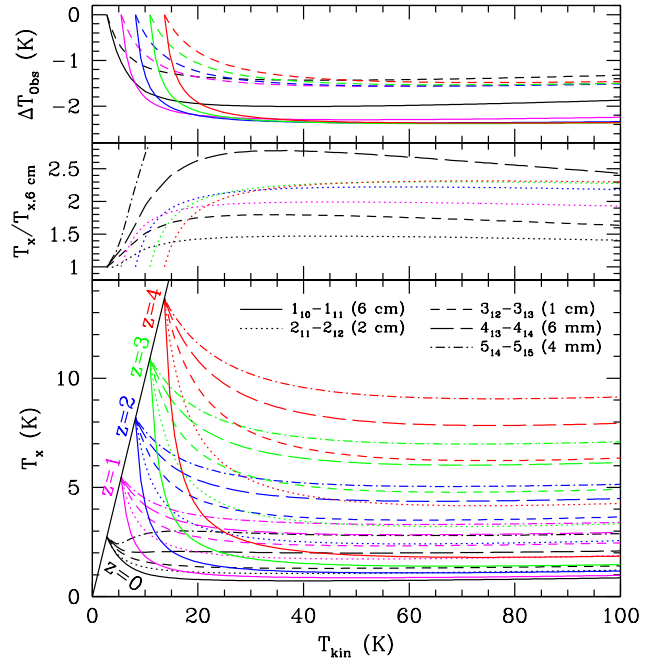}
\caption{Formaldehyde line excitation 
versus gas kinetic temperature.
We plot several redshifts to demonstrate the 
impact of the $(1+z)$ CMB temperature scaling.
Line style indicates transition,
and color indicates redshift.
Bottom:  excitation temperature versus \tkin.  
The black diagonal line shows $T_{\rm x}=T_{\rm kin}$, and
the convergence point of the transitions at each redshift indicates
where radiation, kinetic, and excitation temperatures are equal.  
Middle:  excitation temperature normalized by the 6~cm line excitation temperature versus \tkin.  
We plot all lines at $z=0$ and the 2~cm line at redshifts
$z=0$--4 normalized to the 6~cm line at each redshift.  
Top:  observer-frame temperature decrement $\Delta T_{\rm Obs}$ (Equation \ref{eqn:DTobs}) for
an optically thick line versus \tkin.  We plot redshifts $z=0$--4 for
the representative 6~cm and the 1~cm lines (other lines omitted for clarity).
After the plunge from $T_{\rm kin}=T_{\rm x}=T_{\rm CMB}$, 
\tx\ and $\Delta T_{\rm Obs}$
depend only weakly on \tkin\ with the notable
exception of the higher energy $z=0$ lines, which lie
on the threshold of anti-inversion.
\label{Tx_vs_Tkin}}
\end{figure}

At each redshift, \tkin\ has a $T_{\rm kin}=T_{\rm CMB}=T_{\rm x}$ floor:
the anti-inversion disappears when the radiation, excitation, and gas 
kinetic temperatures are equal (Figure \ref{Tx_vs_Tkin}).
The lower panel of Figure \ref{Tx_vs_Tkin} indicates a nearly redshift-proportional 
line excitation: \tx\ 
becomes insensitive to \tkin\ once collisions dominate the level populations, 
and redshift alone dictates \tx.
Except for $z\sim0$, where \tcmb\ is very low, the line decrements are
insensitive to \tkin\
after an initial drop from $T_{\rm x}=T_{\rm kin}=T_{\rm CMB}$.  
In fact, higher energy $K$-doublet
lines become anti-inverted at $z \gtrsim 1$.  
The 6 and 1~cm  lines show excellent
overlap in $\Delta T_{\rm Obs}$ for $z\geq 1$
and a nearly \tkin-independent trend.  

As the upper panel of Figure \ref{Tx_vs_Tkin} shows, there is little
change in the observed line depth versus redshift or \tkin\ 
with the notable exception of the 6~cm line at $z\simeq0$, and there is 
rapid convergence of all lines at $z\gtrsim1$ for realistic values of \tkin\ ($\gtrsim20$~K).
Figure \ref{Tx_vs_Tkin} also demonstrates that 
there is little variation in the excitation temperature
ratios between lines for $z\gtrsim1$ as a function of gas kinetic temperature.  
Since line ratios indicate the H$_2$ molecular gas density, 
this demonstrates that we do not need precise values for kinetic temperature
to employ the \form\ ``densitometer,''
regardless of redshift and regardless of observed lines.

 For nearly all redshifts and in realistic physical conditions, the 6~cm line shows the 
strongest anti-inversion, and the anti-inversion monotonically decreases
with increasing line energy.  Remarkably, for most scenarios, the 
line decrements grow 
from $z=0$ to $z\sim1$, 
making the lines more detectable at high redshift than locally.  

\section{Discussion}\label{sec:discussion}


Our model predictions for formaldehyde excitation are consistent with observations of molecular clouds in the Galaxy, 
in nearby star-forming galaxies, and in a gravitational lens.  
Galactic dark clouds show a 6~cm line decrement 
of $T_{\rm x}-T_{\rm CMB}\simeq-0.7$ to $-0.4$ K \citep{evans75}, and 
an LVG model applied to Galactic giant molecular clouds obtains comparable line decrements, 
ranging from $-$1 to 0~K \citep{henkel80}.  
The \form\ observations and radiative transfer modeling of star-forming galaxies in the local universe by 
\citet{mangum08} also agree with our models.
The molecular cloud studied toward B0218+357 at $z=0.68$
by \citet{zeiger10} found line decrements of 
$(T_{\rm x}-T_{\rm CMB})/(1+z)\lesssim-2.1$~K for the 6~cm line (for $2\times\ten{3}\pcc<\nh<1\times\ten{4}\pcc$)
and $-$1.5 to $-$1.8~K for the 2~cm transition.

\subsection{A Formaldehyde Deep Field}

An FDF would survey the history of cosmic star formation in a fashion similar
to other star formation proxies such as CO lines, IR/submillimeter luminosity, radio continuum, or UV light.  
But an FDF will differ from traditional proxies in three essential ways:  
(1) line detection is nearly distance-independent and thus a survey for \form\ lines would be a dense 
molecular gas mass-limited survey; 
(2) the \form\ line strength is insensitive to gas kinetic temperature and dust temperature; and
(3) \form\ lines are unaffected by 
dust opacity.
We suspect that an FDF may be as sensitive to extended gas-rich star-forming galaxies such as the
``BzK'' galaxies as it is to compact merger-induced starbursts such as ULIRGs and star formation-dominated
submm galaxies.  While the FDF concept has its shortfalls, such as a reliance on radiative transfer modeling for 
interpretation, it is a useful survey tool that is orthogonal to current star formation survey methods, and
it overcomes the problem of obtaining redshifts for dusty galaxies identified by their dust continuum.
\form\ lines select star-forming gas, eliminating the
uncertain active galactic nucleus dust-heating contribution to continuum star formation proxies.
The strongest signals will arise from the highest molecular gas masses, which will include
submm galaxies, ULIRGs, HyLIRGs, red quasar hosts, dust-obscured galaxies, and BzK galaxies.

The requirements for an FDF are:  
(1) beam-matched observations to the star-forming region(s) in galaxies; 
(2) large bandwidth, which provides a large redshift span; 
(3) spectral line sensitivity; and
(4) either observation of more than one \form\ line or ancillary data 
in other parts of the EM spectrum to break a one-line redshift 
degeneracy.
Radio interferometry meets these criteria and 
captures many galaxies per primary beam, but
reduces the brightness temperature sensitivity. 
One should thus work at higher frequencies and use
a more compact array, but higher frequency 
lines have lower optical depth and higher \tx\  
(reduced $\Delta T_{\rm Obs}$; Figures \ref{Tx_vs_nH2} and 
\ref{Tx_vs_z}).  A compromise can be found, likely using the 2 or 1 cm lines, despite 
the intrinsically stronger 6 cm line signal.

\begin{figure}
\epsscale{1.2}
\plotone{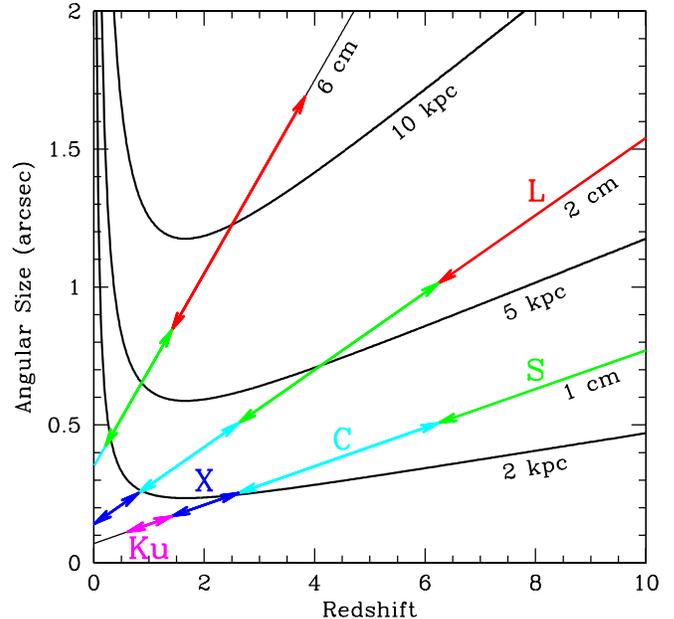}
\caption{Angular sizes of fiducial 2, 5, and 10 kpc molecular gas regions in galaxies
(bold black lines) and the VLA A-array angular resolution 
of the 1, 2, and 6~cm lines of \form\ versus redshift.
Galaxies can be beam-matched if the angular resolution track falls below a given 
physical size locus.  
Colors indicate the VLA wide-band receiver coverage for each line
(L-band spans 1--2~GHz, S spans 2--4~GHz, C spans 4--8~GHz, X spans 8--12~GHz, and Ku spans 12--18~GHz).
\label{angsize}}
\end{figure}

Figure \ref{angsize} shows the angular size of 2, 5, and 10~kpc
molecular regions 
and the Jansky Very Large Array (VLA) synthesized beam 
for the 1, 2, and 6~cm \form\ lines for redshifts 0--10.
The 4~GHz instantaneous frequency coverage of the C-band spans
the remarkable redshift range 
$z=0.8$--6.2 --- most of the epoch of galaxy evolution --- 
because as the 2~cm line redshifts out of the C-band, the 1~cm line
enters the bandpass.  
Even compact star-forming regions in galaxies could be beam-matched
in the X-band over a significant redshift range  for
galaxy evolution studies:  $z=0.2$--0.8 and $z=1.4$--2.6 simultaneously.
A deep observation in one of these bands would detect {\it every} galaxy above the 
gas mass limit in the field of view and would provide spectroscopic redshifts.

\subsection{Formaldehyde Cosmology}

Formaldehyde silhouettes of galaxies against the CMB offer a novel 
probe of cosmology similar to the Sunyaev-Zel'dovich (SZ) effect in 
galaxy clusters.  
The thermal SZ effect depends on the electron density as $n_e$ while the 
X-ray emission from the cluster gas depends on $n_e^2$.  
A comparison 
of the two quantities yields a measure of the line-of-sight length 
scale of the cluster, $\ell$.  If the proper size and line-of-sight size are equal
or can be related via models, then the observed angular size of the cluster yields 
the angular size (or luminosity) distance:
\begin{equation}
  D_A=D_L\,(1+z)^{-2}={\ell\over\theta}.
\end{equation}

A similar measurement may be possible using 
\form\  absorption of the CMB in galaxies.  
\form\ line ratios indicate the local H$_2$ density, and 
line depths yield the \form\ column density.
Since the column density is simply the space density integrated along the line
of sight, 
a comparison of the two quantities measures the line of sight physical size of
the dense gas region of the galaxy.  Assuming the line of sight and 
proper sizes to be equal or the galaxy has a measurable inclination, 
one can measure $D_A$ ($D_L$)
from a spatially resolved spectral line map.

Neither clusters nor galaxies are spherical, but 
many galaxies can be observed per field of view, and in a statistical
aggregate of randomly oriented galaxies, the assumption of equal
size and length scales will be valid. 
This redshift-independent \form\  technique can employ galaxies far beyond the observed
$z\sim 1$ limit on clusters, providing a unique standard ruler from the
end of reionization to the formation of galaxy clusters.

While the number of galaxies
with metal-rich gas at high redshift may decline compared to $z=2$--3, 
molecules detected in quasar hosts up to $z=6.42$ have demonstrated that high 
density regions in the early universe form metals and molecules quickly 
\citep[e.g.,][]{walter2003}.  
This suggests that one may be able to identify standard rulers 
at very high redshift and make unprecedented cosmological measurements 
over a large span of the age of the universe (unlike cosmological 
redshift, $D_L$ and $D_A$ are history-integrated quantities).
While the star-forming regions in galaxies may show size evolution 
with redshift, this method does not rely on a fixed size for galaxies, only on a measurable size.

The largest possible source of uncertainty for \form\ cosmology is the \form\ abundance, which 
is needed to convert from $n({\rm H}_2)$ to $n({\rm\form})$.
Since $\ell\simeq~N({\rm\form})/n({\rm\form})$, but $n({\rm H}_2)=n({\rm \form})/X_{\rm H_2CO}$ is the quantity derived 
from line ratios, the measured size of the molecular region in galaxies is inversely proportional to the
\form\ abundance.


\subsection{Formaldehyde Contamination of \ion{H}{1} 21 cm Tomography at High Redshift}

One significant concern for studies of the \ion{H}{1} 21~cm line before, during, and after the 
EoR is whether \form\ absorption of CMB photons may form a contaminating foreground.
The main lines of interest would be those below or redshifted below
$\sim200$~MHz ($z_{\rm HI}>6$):  the $2_{20}-2_{21}$ 71.1~MHz and 
$3_{21}-3_{22}$ 355.6~MHz lines of para-\form\ are the best
candidates (Figure \ref{H2CO_levels}).  The ground state of the
$K_a=2$ rotation ladder is 
57.6~K above ground, but
radiative transitions to lower para-\form\ states are forbidden.  As discussed in \citet{zeiger10}, warm 
molecular gas can populate the $K_a=2$ states and collisions with H$_2$ will drive these meter-wave transitions 
to very low excitation temperatures, of order 10--30~mK, nearly compensating for the low optical depths of these
lines compared to the centimeter-wave lines.  These low-frequency lines have a negligible excitation 
temperature compared to the CMB such that 
$\Delta T_{\rm Obs}\simeq-(1-e^{-\tau})~2.73$~K~$\simeq-\tau\cdot2.73$~K, and the 
observed line depth is simply determined by its optical depth regardless of redshift.

Comparing \form\ meter-wave optical depths
to cosmological \ion{H}{1} 21 cm signals of nominal brightness temperature 20~mK \citep[e.g.,][]{gnedin04}, we find that the 
$2_{20}-2_{21}$ and $3_{21}-3_{22}$ lines are weaker than the
\ion{H}{1} 21 cm line for $\tau<0.007$.  However, our model predicts 
lines with $\tau\simeq0.2$ and 0.007, respectively.  In 
slightly warmer 60~K gas, the optical depths rise to 0.4 and 0.02, and both lines become a potentially 
problematic source of contamination to \ion{H}{1} 21 cm tomography.  The \ion{H}{1} and \form\ signals may be separable
by angular scale and spectral width, but one can imagine unresolved star formation in foreground large-scale structures
near the peak of cosmic star formation at 
$z\simeq1$--3 creating \form\ 355.6~MHz line signatures that could mimic \ion{H}{1} 21 cm signals at $z\sim10$.


\section{Conclusions}\label{sec:conclusions}

Collisions with H$_2$ cause cm \form\ lines to become ``anti-inverted'' to the point where line
excitation temperatures drop below the rest-frame CMB temperature and thus create silhouettes of the dense
molecular gas regions in galaxies against the uniform background.  We have demonstrated using radiative 
transfer models that cm \form\ lines observed in an FDF
can provide a distance-independent extinction-free mass-limited census of the 
cosmic history of star formation.  The detectability and interpretation of \form\ line surveys is 
nearly independent of redshift and gas kinetic temperature.  We have also examined the possibility that 
\form\ lines can be used to standardize galaxies as cosmic rulers and that anti-inverted meterwave
\form\ lines may confuse or contaminate \ion{H}{1} 21 cm tomography of the EoR.
Significant new observations of \form\ at high redshift will soon be possible using wide-band 
interferometric facilities such as the VLA.

\acknowledgments
We are indebted to K. Eggert for writing support.
We acknowledge the support of the NSF through award GSSP07-0015 from the NRAO and grant AST-0707713.
The LUNAR consortium (\url{http://lunar.colorado.edu}) is funded by the NASA Lunar Science
Institute (Cooperative Agreement NNA09DB30A).

\end{document}